\let\submissionbibliographystyle\bibliographystyle
\def\submissionoriginalstyle{sn-mathphys-ay}
\def\bibliographystyle#1{%
  \def\submissionrequestedstyle{#1}%
  \ifx\submissionrequestedstyle\submissionoriginalstyle
    \submissionbibliographystyle{sn-mathphys-ay-pages}%
  \else
    \submissionbibliographystyle{#1}%
  \fi}
\documentclass[pdflatex,iicol,sn-mathphys-ay]{sn-jnl}
\let\bibliographystyle\submissionbibliographystyle

\usepackage{graphicx}
\usepackage{amsmath,amssymb,amsfonts}
\usepackage{amsthm}
\usepackage{booktabs}
\usepackage{algorithm}
\usepackage{algpseudocode}
\usepackage{placeins}
\usepackage{xcolor}
\newcommand{\reviewblue}[1]{\textcolor{black}{#1}}
\hypersetup{citecolor=black,linkcolor=black,urlcolor=black,filecolor=black}
\newenvironment{reviewmovedfloat}{\begingroup}{\endgroup}

\theoremstyle{thmstyleone}

\theoremstyle{thmstyletwo}

\theoremstyle{thmstylethree}

\newcommand{\SimReadyBaselineReduction}{34.9}

\newcommand{\SimReadyStrongAdvectionReduction}{67.2}
\newcommand{\SimReadyZeroAdvectionReduction}{3.8}

\newcommand{\SimReadyJointTable}{%
\begin{table*}[t]
\caption{{Mat\'ern results} from 120 matched simulations per scenario at budget $(6,4,3)$. Both rules converged on all replicates. RMSE is the advection-vector RMSE in $10^{-3}$ degrees per hour. Reduction is $100(1-\mathrm{RMSE}_{A}/\mathrm{RMSE}_{T})$, where $A$ and $T$ denote advection-aligned and target-centered conditioning. Time is mean graph-construction plus optimization time, excluding shared initialization and I/O.}
\label{tab:simready-joint}
\centering\small
\setlength{\tabcolsep}{3.5pt}
\begin{tabular}{lrrrrr}
\toprule
Scenario & \shortstack{Target-\\centered\\RMSE} & \shortstack{Advection-\\aligned\\RMSE} & \shortstack{Reduction ($\%$)} & \shortstack{Target-\\centered\\time (s)} & \shortstack{Advection-\\aligned\\time (s)} \\
\midrule
Reference range $(0.2,0.3)$ & 3.371 & 2.194 & 34.9 & 9.36 & 9.99 \\
Short range $(0.1,0.15)$ & 1.255 & 0.768 & 38.7 & 17.90 & 19.88 \\
Long range $(0.4,0.6)$ & 11.370 & 8.432 & 25.8 & 24.11 & 11.99 \\
Weak advection & 2.366 & 2.118 & 10.5 & 9.76 & 10.04 \\
Strong advection & 7.294 & 2.395 & 67.2 & 7.83 & 9.78 \\
Zero advection & 2.367 & 2.278 & 3.8 & 9.75 & 9.83 \\
\bottomrule
\end{tabular}
\end{table*}%
}

\begin{document}

\title[Advection-aligned block Vecchia]{Advection-aligned block Vecchia inference for spatio-temporal Gaussian processes}

\author*[1]{\fnm{Joonwon} \sur{Lee}}
\email{{\color{black}joonwon.lee22@gmail.com}}
\author[1]{\fnm{Michael L.} \sur{Stein}}

\affil*[1]{\orgname{Rutgers University},
  \orgaddress{\city{Piscataway}, \state{NJ}, \country{United States}}}

\abstract{%

Vecchia approximations enable scalable likelihood-based inference for large Gaussian process models, but their accuracy depends on the choice of conditioning sets. For spatio-temporal processes with advection, target-centered conditioning can omit informative past observations. {Motivated by total column ozone observations from the Geostationary Environment Monitoring Spectrometer (GEMS)}, we propose a data-driven, advection-aligned block Vecchia approximation that incorporates estimated transport direction into temporal conditioning geometry. {Observation locations need not lie on a regular grid.} An initial advection vector is estimated from the data and used once to select transport-aligned past-time conditioning blocks. The resulting graph is held fixed during optimization, while covariance and advection parameters remain free to vary. This separates neighbor selection from covariance estimation, avoids repeated parameter-dependent neighbor searches, and provides an interpretable connection between conditioning geometry and estimated transport. Batched graphics processing unit (GPU) operations evaluate the likelihood and its gradients. {In a matched comparison of daily GEMS datasets from July 2024 and July 2025, advection-aligned conditioning attained a lower fitted Vecchia negative log-likelihood than target-centered conditioning on 53 of 61 days. In six Mat\'ern simulation scenarios with 120 matched eight-hour fields each, advection-aligned conditioning reduced advection-vector root mean squared error relative to target-centered conditioning at the same conditioning-block budget, with the largest reduction under strong transport.}
}

\keywords{Vecchia approximation, Gaussian processes, spatio-temporal statistics, {total column ozone}, scalable computation}


\maketitle

\section{Introduction}\label{sec:introduction}

Exact likelihood-based inference with a dense Gaussian process covariance matrix generally requires \(O(n^3)\) floating-point operations and \(O(n^2)\) memory, limiting its use for large spatial and spatio-temporal datasets.

\begingroup
\color{black}
\hypersetup{citecolor=black,linkcolor=black}
Let \(\boldsymbol{y}=(y_1,\ldots,y_n)^\top\) denote the observations
and let \(X\in\mathbb{R}^{n\times p}\) be a design matrix. We
consider the Gaussian model
\begin{equation}
\boldsymbol{y}
\sim
N_n\bigl(X\beta,\Sigma_{\theta}\bigr),
\label{eq:gp-model}
\end{equation}
where \(\beta\in\mathbb{R}^{p}\) contains the mean parameters and
\(\theta\) contains the covariance parameters. We assume that
\(\Sigma_{\theta}\) is positive definite for the parameter values
under consideration.

For any ordering of the observations, the joint density has the
exact factorization
\begin{equation}
\begin{aligned}
p_{\theta,\beta}(\boldsymbol{y})
&=
p_{\theta,\beta}(y_1)\\[-1pt]
&\quad{}\times
\prod_{j=2}^{n}
p_{\theta,\beta}(y_j\mid y_1,\ldots,y_{j-1}).
\end{aligned}
\label{eq:exact-factorization}
\end{equation}
The Vecchia approximation \citep{vecchia1988,katzfussguinness2021}
comes from replacing the complete
preceding history in each conditional with
\(\boldsymbol y_{c(j)}\):
\begin{equation}
\begin{aligned}
\widetilde p_{\theta,\beta}(\boldsymbol{y})
&=
p_{\theta,\beta}(y_1)\\[-1pt]
&\quad{}\times
\prod_{j=2}^{n}
p_{\theta,\beta}(y_j\mid\boldsymbol{y}_{c(j)}),
\end{aligned}
\label{eq:vecchia}
\end{equation}
where \(c(j)\subseteq\{1,\ldots,j-1\}\), \(|c(j)|\leq m\),
and \(m\) denotes the maximum conditioning-set size. Here,
\(\boldsymbol{y}_{c(j)}\) contains the observations indexed
by \(c(j)\).

Each conditional density in \eqref{eq:vecchia} is evaluated under the Gaussian model
in \eqref{eq:gp-model}.
\par
\endgroup

\begingroup
\color{black}
\hypersetup{citecolor=black,linkcolor=black}
For single-observation targets with at most \(m\) conditioning observations, evaluating this approximation costs \(O(nm^3)\), with \(m\ll n\). Approximation quality depends on the choice of conditioning sets.
\par
\endgroup

\begingroup
\color{black}
\hypersetup{citecolor=black,linkcolor=black}
In spatial applications, nearby predecessors are often selected because
more distant observations may add little information about the target
once nearby observations are included---the spatial screening effect
\citep{stein1999book}.
With advection, however, informative past observations may lie near a
transported location rather than the current target. A target-centered
conditioning set can therefore omit these observations even when the
covariance model includes advection. We estimate a reference advection
vector \(\widetilde{\boldsymbol v}\) from adjacent hourly fields. For a target
block with centroid \(\boldsymbol s\), we center the past-block search at
\(\boldsymbol s-u\widetilde{\boldsymbol v}\), rather than at
\(\boldsymbol s\), for one- and two-hour lags \(u\). We call this
construction advection-aligned.
\par
\endgroup

Computational developments have followed two complementary directions:
reducing or reusing the local calculations required by the approximation,
and evaluating many local systems concurrently using GPU parallelism.

The first direction seeks to reduce or reuse distinct local matrix
factorizations. Stein, Chi, and Welty~\citeyearpar{stein2004} extended
Vecchia's construction to block prediction vectors. {When target blocks have sizes proportional to \(m\), there are \(O(n/m)\) local factorizations, each costing \(O(m^3)\), yielding \(O(nm^2)\) total work.} They also noted that overlap
among conditioning sets could be exploited to reduce computation.
Guinness~\citeyearpar{guinness2018} later developed this direction into an
automatic grouping method applied after the ordering and neighbor sets
have been selected. Grouping can improve approximation accuracy by
enlarging the effective conditioning sets. It also reduces repeated local
factorizations by evaluating several conditional contributions from a
single covariance factorization per group.
Guinness~\citeyearpar{guinness2018} also studied ordering strategies,
including maximum-minimum-distance ordering. These block and grouping
constructions were particularly well suited to CPU-based implementations,
where reducing repeated local linear algebra could substantially lower
the total computational cost.
{Automatic grouping can produce local systems of different sizes, complicating
fixed-size GPU batching. Padding adds redundant computation, while size-specific
batching divides the workload into smaller batches. These adjustments can reduce
parallel efficiency.}

{A second direction exploits GPU parallelism across independent
local Vecchia conditionals. \citet{pan2024gpu} used batched Cholesky
factorizations and triangular solves to compute local log-determinant and
quadratic-form contributions, which are summed to evaluate the Vecchia
likelihood. \citet{jamesguinness2024} compared thread-per-observation,
block-per-observation, and batched GPU implementations of scalar Vecchia
factors. \citet{pan2025block} developed a GPU implementation of block
Vecchia using K-means partitions and variable-size batching, and
\citet{panSBV} extended batched evaluation to distributed Scaled Block
Vecchia. Computational reuse and GPU parallelism can be combined. We use
batched Cholesky factorizations and triangular solves for block targets.}

The spatial nearest-neighbor Gaussian process (NNGP) of \citet{datta2016nngp} extends Vecchia's ordered conditional construction to a valid spatial process for scalable hierarchical modeling and uses Euclidean-nearest predecessors in its standard form. The dynamic spatio-temporal NNGP of \citet{datta2016dnngp} instead ranked predecessors by the parent covariance within predetermined space--time histories and proposed adaptive searches over parameter-free eligible sets. {\citet{jurek2022hierarchical} developed a hierarchical Vecchia approximation for scalable spatio-temporal filtering.} \citet{kang2023correlation} used the correlation distance \(\tau_C(i,j)=\{1-|\rho_{ij}|\}^{1/2}\) for both nearest-neighbor conditioning and maximum--minimum ordering.

\citet{gyger2026weather} developed GPU-accelerated matrix computations and correlation-based neighbor searches for nonseparable spatio-temporal Gaussian process models, updating conditioning sets during likelihood optimization. Our focus is a fixed, advection-aligned construction of past-time conditioning blocks, combined with GPU-batched evaluation of the resulting block conditional likelihoods.

The conditioning indices are selected once and held fixed throughout
optimization. Covariance matrices and their derivatives are recomputed
using the current parameters, but neighbor searches are not repeated.
{We group the local conditional likelihood terms according to the available time layers: the current hour only, the current hour and the one-hour lag, or the current hour and the one- and two-hour lags. Within each group, local systems with matching conditioning and target dimensions are stacked into tensors for batched Cholesky factorizations and triangular solves on the GPU.}
We compare advection-aligned and target-centered conditioning at the same
block budget, assessing parameter estimation in simulations and fitted
Vecchia likelihoods in both simulations and GEMS data.
The empirical initializer requires no preliminary
covariance likelihood optimization and can also supply the advection
starting value for correlation-based neighbor selection.
\reviewblue{The lattice-based construction is best suited to observations
that are not too unevenly spaced. The GEMS ozone observations described in
Section~\ref{sec:gems-data} lie on a nearly regular grid, as do observations in
many satellite-based environmental datasets.}

\section{{Block Vecchia likelihood computation}}\label{sec:background}

\begingroup
\color{black}
\subsection{{Block conditionals and Cholesky whitening}}
\label{sec:block-conditional}
\endgroup

Let \(T_1,\ldots,T_M\) be ordered, nonempty target sets partitioning the
observations:
\begin{equation}
\bigcup_{i=1}^M T_i=\{1,\ldots,n\},\qquad
T_i\cap T_j=\varnothing\quad(i\ne j).
\label{eq:target-partition}
\end{equation}
For block \(i\), the conditioning set satisfies
\begin{equation}
C_i\subseteq\bigcup_{j<i}T_j,\qquad |C_i|\leq m.
\label{eq:block-conditioning}
\end{equation}
A block index includes time when blocks are formed separately at each time.
The block Vecchia working density is
\begin{equation}
\widetilde p_{\theta,\beta}(\boldsymbol y)
=\prod_{i=1}^M p_{\theta,\beta}
(\boldsymbol y_{i,T}\mid\boldsymbol y_{i,C}),
\label{eq:block-vecchia}
\end{equation}
where the subvectors contain observations in \(T_i\) and \(C_i\).
An empty \(C_i\) gives a marginal factor. The partition and predecessor
conditions ensure a joint density for a fixed graph; singleton targets
recover scalar Vecchia.

Place conditioning observations before targets in each local system:
\begin{equation}
\boldsymbol y_i=
\begin{pmatrix}\boldsymbol y_{i,C}\\\boldsymbol y_{i,T}\end{pmatrix},
\qquad
X_i=\begin{pmatrix}X_{i,C}\\X_{i,T}\end{pmatrix},
\label{eq:local-data}
\end{equation}
where \(X_{i,C}\) and \(X_{i,T}\) are rows of the global design matrix.
The local covariance and its Cholesky factor have the form
\begin{equation}
\begin{aligned}
\Sigma_i&=
\begin{pmatrix}
\Sigma_{CC,i}&\Sigma_{CT,i}\\
\Sigma_{TC,i}&\Sigma_{TT,i}
\end{pmatrix}=L_iL_i^\top,\\
L_i&=\begin{pmatrix}
L_{CC,i}&0\\L_{TC,i}&L_{TT,i}
\end{pmatrix}.
\end{aligned}
\label{eq:local-cholesky}
\end{equation}
These matrices depend on \(\theta\), which is suppressed in the local
notation for simplicity.

The conditional regression matrix and conditional covariance are
\begin{align}
B_i
&=
\Sigma_{TC,i}\Sigma_{CC,i}^{-1}
=
L_{TC,i}L_{CC,i}^{-1},
\label{eq:conditional-regression}\\
\Sigma_{T\mid C,i}
&=
\Sigma_{TT,i}
-
\Sigma_{TC,i}\Sigma_{CC,i}^{-1}\Sigma_{CT,i}
=
L_{TT,i}L_{TT,i}^{\top}.
\label{eq:conditional-covariance}
\end{align}
{Consequently,}
\begin{equation}
\begin{aligned}
\boldsymbol{\mu}_{i\mid C}(\beta)
&=
X_{i,T}\beta\\[-1pt]
&\quad{}+
B_i(\boldsymbol{y}_{i,C}-X_{i,C}\beta),\\[2pt]
\boldsymbol{y}_{i,T}\mid\boldsymbol{y}_{i,C}
&\sim
{N}\!\left(
\boldsymbol{\mu}_{i\mid C}(\beta),
\Sigma_{T\mid C,i}
\right).
\end{aligned}
\label{eq:conditional-normal}
\end{equation}
{Write \(a_i(\beta)=\boldsymbol y_{i,T}-\boldsymbol\mu_{i\mid C}(\beta)\)
for the conditional residual. When \(C_i\) is empty, terms involving
\(B_i\) are omitted.}

The conditional residual can be whitened directly through the local
Cholesky factor. Define the target rows of the triangular solves by
\begin{equation}
U_{y,i}=[L_i^{-1}\boldsymbol y_i]_{T_i},\qquad
U_{x,i}=[L_i^{-1}X_i]_{T_i}.
\label{eq:target-whitening}
\end{equation}
Here \([\cdot]_{T_i}\) selects the last \(d_i=|T_i|\) rows under
\eqref{eq:local-data}. {Since conditioning rows precede target rows,}
\begingroup
\color{black}
\begin{equation}
\begin{aligned}
U_{y,i}-U_{x,i}\beta
&=L_{TT,i}^{-1}a_i(\beta),\\
a_i(\beta)^\top\Sigma_{T\mid C,i}^{-1}a_i(\beta)
&=\left\|U_{y,i}-U_{x,i}\beta\right\|^2.
\end{aligned}
\label{eq:conditional-quadratic}
\end{equation}
\endgroup
{In implementation, we solve \(L_iU_i=[X_i,\boldsymbol y_i]\)
and retain only the last \(d_i\) rows, corresponding to the target block.
These rows provide the conditional quadratic term without explicitly forming \(B_i\) or
\(\Sigma_{T\mid C,i}^{-1}\).}

\begingroup
\color{black}
\subsection{{Profile likelihood and GPU batching}}
\label{sec:batched-likelihood}
\endgroup

The conditional log-determinant is also available from the target block
of the Cholesky factor:
\begin{equation}
\log\det\Sigma_{T\mid C,i}
=
2\sum_{k=1}^{d_i}\log(L_{TT,i})_{kk}.
\label{eq:conditional-logdet}
\end{equation}
Combining \eqref{eq:conditional-quadratic} and
\eqref{eq:conditional-logdet}, the block Vecchia log-likelihood is
\begin{equation}
\begin{aligned}
\widetilde\ell(\theta,\beta)=-\frac12\Biggl[
&n\log(2\pi)
+2\sum_{i=1}^M\sum_{k=1}^{d_i}\log(L_{TT,i})_{kk}\\
&+\sum_{i=1}^M\|U_{y,i}-U_{x,i}\beta\|^2\Biggr],
\end{aligned}
\label{eq:vecchia-loglikelihood}
\end{equation}
with \(n=\sum_i d_i\). Only target conditional terms are included;
using the full local determinant would also count the density of the
conditioning observations and would not yield \eqref{eq:block-vecchia}.

For each \(\theta\), accumulate
\begin{equation}
\begin{aligned}
H(\theta)&=\sum_{i=1}^M U_{x,i}^\top U_{x,i},\\[2pt]
g(\theta)&=\sum_{i=1}^M U_{x,i}^\top U_{y,i},\\[2pt]
s(\theta)&=\sum_{i=1}^M U_{y,i}^\top U_{y,i},\\[2pt]
d(\theta)&=2\sum_{i=1}^M\sum_{k=1}^{d_i}\log(L_{TT,i})_{kk}.
\end{aligned}
\label{eq:likelihood-summaries}
\end{equation}
If \(H(\theta)\) is nonsingular, the generalized least squares (GLS) mean estimate is
\begin{equation}
\widehat\beta(\theta)=H(\theta)^{-1}g(\theta),
\label{eq:beta-estimate}
\end{equation}
and the profile log-likelihood is
\begin{equation}
\begin{aligned}
\widetilde\ell_{\mathrm{prof}}(\theta)
&=\widetilde\ell(\theta,\widehat\beta(\theta))\\[2pt]
&=-\frac12\Bigl[n\log(2\pi)+d(\theta)+s(\theta)\\[-1pt]
&\qquad{}-g(\theta)^\top H(\theta)^{-1}g(\theta)\Bigr].
\end{aligned}
\label{eq:profile-loglikelihood}
\end{equation}
The same mean coefficient vector is shared across blocks and recomputed
for each covariance parameter vector.
{The GLS coefficients are computed by solving
\(H(\theta)\widehat{\beta}(\theta)=g(\theta)\) using a Cholesky
factorization of \(H(\theta)\).}

Although the working density is defined sequentially, its local
contributions for observed data can be evaluated independently at a fixed
parameter value. The local covariance matrices, Cholesky factorizations,
and triangular solves can therefore be processed using GPU-accelerated
batched computations \citep{pan2024gpu,jamesguinness2024}.
For example, \(b\) local systems with the same conditioning and target
sizes can be stored as a \(b\times q\times q\) covariance array, where
\(q=|C_i|+|T_i|\). The corresponding responses and design matrices can be
combined into a \(b\times q\times(p+1)\) array of right-hand sides.
Systems with different sizes can be processed in separate batches.

After the batched triangular solves, the target rows are extracted and
their contributions to \(H(\theta)\), \(g(\theta)\), \(s(\theta)\), and
\(d(\theta)\) are accumulated. This avoids forming the global
\(n\times n\) covariance factorization or storing all transformed
observations at once. Here, target blocks define the conditional factors,
whereas GPU batches specify which local matrix operations are executed
together. GPU batching does not require the local neighborhood geometries
or covariance matrices to be identical. A separate Cholesky factor is computed for each local
covariance matrix, while many such decompositions and triangular solves
are executed concurrently on the GPU. This is useful when irregular
observation patterns or advection-aligned conditioning produce different
local geometries across target blocks.

\section{{GEMS total column ozone data}}
\label{sec:gems-data}

{We analyze total column ozone (TCO) from the Geostationary Environment
Monitoring Spectrometer (GEMS; \citealp{kim2020gems}) over
\(3^\circ\mathrm S\)--\(2^\circ\mathrm N\) and
\(121^\circ\mathrm E\)--\(131^\circ\mathrm E\) during July 2024 and July 2025.
Each daily dataset contains eight consecutive hourly scenes, with nominal
scan-start times from 00:45 to 07:45 UTC. July 24, 2025, is excluded because
one hourly field is unavailable, leaving 31 days from 2024 and 30 days from
2025. We analyze the 61 days separately because the daytime observation
sequences are separated by overnight gaps.}

{The observation coordinates are approximately regular but exhibit
spatial irregularities and small hour-to-hour shifts. We associate observations
with a reference lattice having latitude and longitude spacings of
\(0.044^\circ\) and \(0.063^\circ\), approximately \(4.87\) km north--south and
\(7.00\)--\(7.01\) km east--west. The latitude spacing was chosen to accommodate
the greater irregularity in the reported latitude coordinates. Matching uses
an hour-adjusted lattice and requires absolute latitude and longitude
discrepancies no greater than \(0.022^\circ\) and \(0.0315^\circ\), respectively.
We retain at most one observation per cell; unmatched cells remain missing
and are not imputed.}

{The reference lattice supports the block partition, max--min
ordering, and fast Fourier transform (FFT)-based initialization of advection\reviewblue{,
as described in Section~{\hypersetup{linkcolor=black}\ref{sec:methodology}}}.
Covariance matrices, however, are evaluated at the retained observations'
original source coordinates rather than the reference-lattice coordinates.}

\section{{Advection-aligned block Vecchia approximation}}\label{sec:methodology}

\subsection{{Advection-adjusted spatial lag}}
\label{sec:shifted-lag}

Atmospheric transport can shift the spatial location of cross-time ozone dependence \citep{strahan2011}. Under a constant linear-advection approximation with velocity \(\boldsymbol{v}\), we retain the original observation coordinates and represent transport through the advection-adjusted spatial lag
\[
\boldsymbol{h}-\boldsymbol{v}u.
\]
Here \(u>0\) denotes the later time minus the earlier time, and \(\boldsymbol{h}\) is the corresponding later-location minus earlier-location spatial lag. Hence, candidate observations at temporal lag \(u\) may be relevant near \(\boldsymbol{h}=\boldsymbol{v}u\), rather than only near \(\boldsymbol{h}=\boldsymbol{0}\).

\subsection{{Masked-FFT advection initialization}}
\label{sec:advection-initialization}

Let \(Z_t(\boldsymbol{x})\) denote the field at time \(t\), represented on a
regular reference lattice with spacings \(\Delta_1\) and \(\Delta_2\), and
let \(M_t(\boldsymbol{x})\in\{0,1\}\) be the corresponding observation mask.

{To motivate the initializer, consider the stationary mean-zero
field \(W_t(\boldsymbol{x})=Z_t(\boldsymbol{x})-\mathbb E[Z_t(\boldsymbol{x})]\)
with the advection-adjusted covariance used below:}
\begingroup
\color{black}
\begin{equation}
\begin{aligned}
d_{\theta}(\boldsymbol h,u)
&=\left\{\sum_{r=1}^2\frac{(h_r-v_ru)^2}{\rho_r^2}
+\frac{u^2}{\rho_u^2}\right\}^{1/2},\\
C_{\theta}(\boldsymbol h,u)
&=c_{\theta}\bigl(d_{\theta}(\boldsymbol h,u)\bigr).
\end{aligned}
\label{eq:advected-covariance}
\end{equation}
\endgroup
{Here \(\rho_1,\rho_2,\rho_u>0\) are range parameters and
\(c_{\theta}\) decreases with distance.}
\reviewblue{For fixed \(u>0\), \(d_{\theta}(\boldsymbol h,u)\) is minimized
at \(\boldsymbol h=\boldsymbol vu\), where the physical distance between
the target location and the past location, after accounting for transport,
is zero.}

For each nonempty time slice, let \(\overline Z_t\) denote the spatial mean over the observed locations and define the zero-filled centered array by
\begin{equation}
\begin{aligned}
\overline Z_t
&=
\frac{\sum_{\boldsymbol{x}}M_t(\boldsymbol{x})Z_t(\boldsymbol{x})}
     {\sum_{\boldsymbol{x}}M_t(\boldsymbol{x})},
\\[3pt]
Y_t(\boldsymbol{x})
&=
\begin{cases}
Z_t(\boldsymbol{x})-\overline Z_t,
    & M_t(\boldsymbol{x})=1,\\
0,  & M_t(\boldsymbol{x})=0.
\end{cases}
\end{aligned}
\label{eq:initializer-centering}
\end{equation}

{For adjacent time slots separated by \(\Delta u\), the
population cross-time semivariogram is}
\begingroup
\color{black}
\[
\begin{aligned}
\gamma_1(\boldsymbol{h})
&=\frac{1}{2}\mathbb{E}\Bigl[
\bigl\{W_{t+1}(\boldsymbol{x}+\boldsymbol{h})
-W_t(\boldsymbol{x})\bigr\}^2\Bigr]\\
&=C_{\theta}(\boldsymbol 0,0)-C_{\theta}(\boldsymbol h,\Delta u).
\end{aligned}
\]
\endgroup
{Thus its population minimum occurs at
\(\boldsymbol h=\boldsymbol v\Delta u\). The initializer uses a
sample-centered, valid-pair analogue, pooling adjacent-time transitions.
Masks exclude missing pairs, so zero filling does not add observations.
Define the unnormalized discrete cross-correlation}
\begin{equation}
R_{fg}(\boldsymbol{h})
=
\sum_{\boldsymbol{x}}
f(\boldsymbol{x}+\boldsymbol{h})g(\boldsymbol{x}).
\label{eq:correlation-convention}
\end{equation}
For each adjacent-time pair, compute
\begin{equation}
\begin{aligned}
N_t(\boldsymbol{h})
&=R_{M_{t+1},M_t}(\boldsymbol{h}),\\
A_t(\boldsymbol{h})
&=R_{M_{t+1},Y_t^2}(\boldsymbol{h}),\\
B_t(\boldsymbol{h})
&=R_{Y_{t+1}^2,M_t}(\boldsymbol{h}),\\
C_t(\boldsymbol{h})
&=R_{Y_{t+1},Y_t}(\boldsymbol{h}).
\label{eq:initializer-correlations}
\end{aligned}
\end{equation}
Here, \(N_t(\boldsymbol{h})\) counts the valid observation pairs, while \(A_t(\boldsymbol{h})+B_t(\boldsymbol{h})-2C_t(\boldsymbol{h})\) is the corresponding sum of squared differences. Pooling these quantities over adjacent-time transitions gives
\begin{equation}
\widehat\gamma_1(\boldsymbol{h})
=
\frac{
\sum_t\{A_t(\boldsymbol{h})+B_t(\boldsymbol{h})-2C_t(\boldsymbol{h})\}
}{
2\sum_tN_t(\boldsymbol{h})
}.
\label{eq:pooled-one-step-semivariogram}
\end{equation}
At each lag, transitions are weighted by their numbers of valid pairs.

{The four cross-correlations are evaluated using zero-padded
FFTs. With \(N_{\mathrm{grid}}\) reference-lattice cells, evaluating all
spatial lags requires \(O(N_{\mathrm{grid}}\log N_{\mathrm{grid}})\)
operations per adjacent-time pair, compared with
\(O(N_{\mathrm{grid}}^2)\) for direct calculation. The masks restrict the
calculation to observed pairs \citep{marcotte1996fft,lauzonhorning2025}.
Padding requirements are given in
Appendix~\ref{app:masked-fft-implementation}.}

Let \(\mathcal H\) denote a prespecified set of candidate spatial lags and define
\begin{equation}
\mathcal H_{\mathrm{valid}}
=
\left\{
\boldsymbol{h}\in\mathcal H:
\sum_tN_t(\boldsymbol{h})\geq N_{\min}
\right\}.
\end{equation}
{We smooth the retained values with a normalized Gaussian
kernel of bandwidth \(h_{\mathrm{smooth}}\), excluding missing lag values
while permitting interpolation. The finite smoothed minimum is refined
by a local quadratic fit when its safeguard criteria hold; otherwise,
the discrete minimum is retained
(Appendix~\ref{app:masked-fft-implementation}). If no smoothed value is
finite, initialization fails.}

{Let \(\widehat{\boldsymbol k}\) denote the selected lag in
lattice-cell units and \(D=\operatorname{diag}(\Delta_1,\Delta_2)\).
Converting displacement to velocity gives the reference vector}
\begin{equation}
\widetilde{\boldsymbol v}
=
\frac{D\widehat{\boldsymbol k}}{\Delta u}.
\label{eq:grid-advection-anchor}
\end{equation}
The components of \(\widetilde{\boldsymbol v}\) therefore have units of spatial coordinate per unit time.

{The empirical minimum provides a graph-construction reference,
not a final estimate of physical advection. This step does not fit
variance, range, or covariance-shape parameters.}

\begin{algorithm}[htbp]
\caption{Masked-FFT initialization}
\label{alg:masked-fft-initializer}
\begin{algorithmic}[1]
\Require Gridded fields and masks \(\{Z_t,M_t\}\), lattice spacing
\(D\), {time spacing \(\Delta u\)}, candidate-lag set \(\mathcal H\), threshold \(N_{\min}\), bandwidth
\({h_{\mathrm{smooth}}}\), and Hessian condition-number bound \(\kappa_{\max}\).
\Ensure Reference vector \(\widetilde{\boldsymbol v}\).

\State Center each nonempty time slice over its observed locations and
zero-fill the missing cells.
\State For all adjacent-time pairs, compute and pool
\(N_t,A_t,B_t,C_t\) using zero-padded FFT correlations.
\State Evaluate \(\widehat\gamma_1\), exclude lags supported by fewer than
\(N_{\min}\) pairs, and apply normalized Gaussian smoothing.
\State Find the finite smoothed minimum and apply the safeguarded
\(3\times3\) quadratic refinement.
\State Convert the refined displacement using
\eqref{eq:grid-advection-anchor}.
\end{algorithmic}
\end{algorithm}

\subsection{{Conditioning graph construction}}
\label{sec:advection-conditioning}

The data-driven reference vector \(\widetilde{\boldsymbol v}\) obtained in
Section~\ref{sec:advection-initialization} is used to select the proposed
block-level conditioning indices and supplies the starting value of the
covariance advection parameter in both fits considered below. The selected
indices define a sparse graph that is held fixed throughout likelihood
optimization, but the covariance advection parameter \(\boldsymbol v\) is not
held fixed. It remains an optimization parameter and, together with the other
current covariance parameters, determines every local covariance matrix.
Likelihood evaluation treats the realized initializer and graph indices as
fixed; it does not model their selection from the observations.

\reviewblue{In a pilot test, after the first L-BFGS call (up to 20 iterations),
we rebuilt the past-time conditioning sets once using the current advection
estimate. We then continued optimization with the updated sets held fixed.
This update increased runtime, mainly through additional likelihood-gradient
evaluations, without reducing the fitted Vecchia negative log-likelihood (NLL).}

Let \(\{\boldsymbol g_r:r=1,\ldots,G\}\) denote the reference lattice used for
graph construction and let \(\mathcal O_t\) contain the cells observed at time
\(t\).  Partition the lattice into nonoverlapping rectangular blocks
\(\mathcal B_1,\ldots,\mathcal B_J\), with centroids
\begin{equation}
\boldsymbol c_j
=
\frac{1}{|\mathcal B_j|}
\sum_{r\in\mathcal B_j}\boldsymbol g_r.
\end{equation}
The centroids are put in max--min order \citep{guinness2018}: the first is
chosen closest to their coordinatewise mean, and each subsequent centroid
maximizes its minimum Euclidean distance from those already selected. The
blocks are relabeled in this order.

At time \(t\), block \(j\) has the multivariate target
\begin{equation}
\boldsymbol y_{t,j}
=
\bigl(y_t(\boldsymbol s_{t,r}):
r\in\mathcal B_j\cap\mathcal O_t\bigr),
\end{equation}
where \(\boldsymbol s_{t,r}\) is the retained source location assigned to cell
\(r\). For irregularly located observations, reference lattice coordinates
determine block membership, ordering, and conditioning indices, whereas each
local covariance matrix is evaluated at the retained original source
coordinates. Empty target blocks contribute no conditional factor.

Let \(m_0\), \(m_1\), and \(m_2\) be the maximum numbers of conditioning
blocks selected at times \(t\), \(t-1\), and \(t-2\), respectively. At lag
zero, admissible blocks must precede target block \(j\) in the max--min
ordering. We therefore let \(\mathcal N_j^{(0)}\) contain up to \(m_0\)
nearest predecessor blocks:
\begin{equation}
\mathcal N_j^{(0)}\subseteq\{1,\ldots,j-1\},
\qquad
|\mathcal N_j^{(0)}|\leq m_0.
\end{equation}
This predecessor restriction is required only for blocks at the target time;
all blocks at earlier times already precede the target in the temporal order.

Under the advection-adjusted lag convention
\(\boldsymbol h-\boldsymbol v u\), a current target is associated with past
locations in the direction \(-\widetilde{\boldsymbol v}\). Let \(\Delta u\)
be the separation between adjacent time slots. The \emph{advection-aligned}
construction uses the query centers
\begin{equation}
\boldsymbol q_{j,\ell}
=\boldsymbol c_j-\ell\widetilde{\boldsymbol v}\Delta u,
\qquad \ell\in\{1,2\}.
\label{eq:nominal-query-centres}
\end{equation}
Let \(\operatorname{NN}_m(\boldsymbol q;\mathcal A)\) denote the indices of
up to \(m\) block centroids nearest in Euclidean distance to
\(\boldsymbol q\), among the candidate indices \(\mathcal A\), using a
fixed rule to resolve equal distances. With \(\mathcal A=\{1,\ldots,J\}\),
the temporal neighbors are
\begin{equation}
\mathcal N_j^{(\ell)}
=\operatorname{NN}_{m_\ell}(\boldsymbol q_{j,\ell};\mathcal A),
\qquad \ell\in\{1,2\}.
\label{eq:nominal-temporal-neighbours}
\end{equation}
Thus the one- and two-step searches are centered at the transport locations
implied by the initial advection estimate. Near the domain boundary, the nearest available blocks
are selected under the same rule; the query center need not lie inside the
domain.

The \emph{target-centered} comparison replaces both query centers by
\(\boldsymbol c_j\). Its past-time searches therefore use the same spatial
distance ranking at both lags, truncated to the respective budgets
\(m_1\) and \(m_2\). {The same-time predecessor search is unchanged; only the
past-time query centers differ. Both optimizations start from the same
daily \(\widetilde{\boldsymbol v}\) and estimate \(\boldsymbol v\) freely:
target-centered conditioning does not fix covariance advection to zero.}

\begin{figure*}[!t]
\centering
\includegraphics[width=\textwidth]{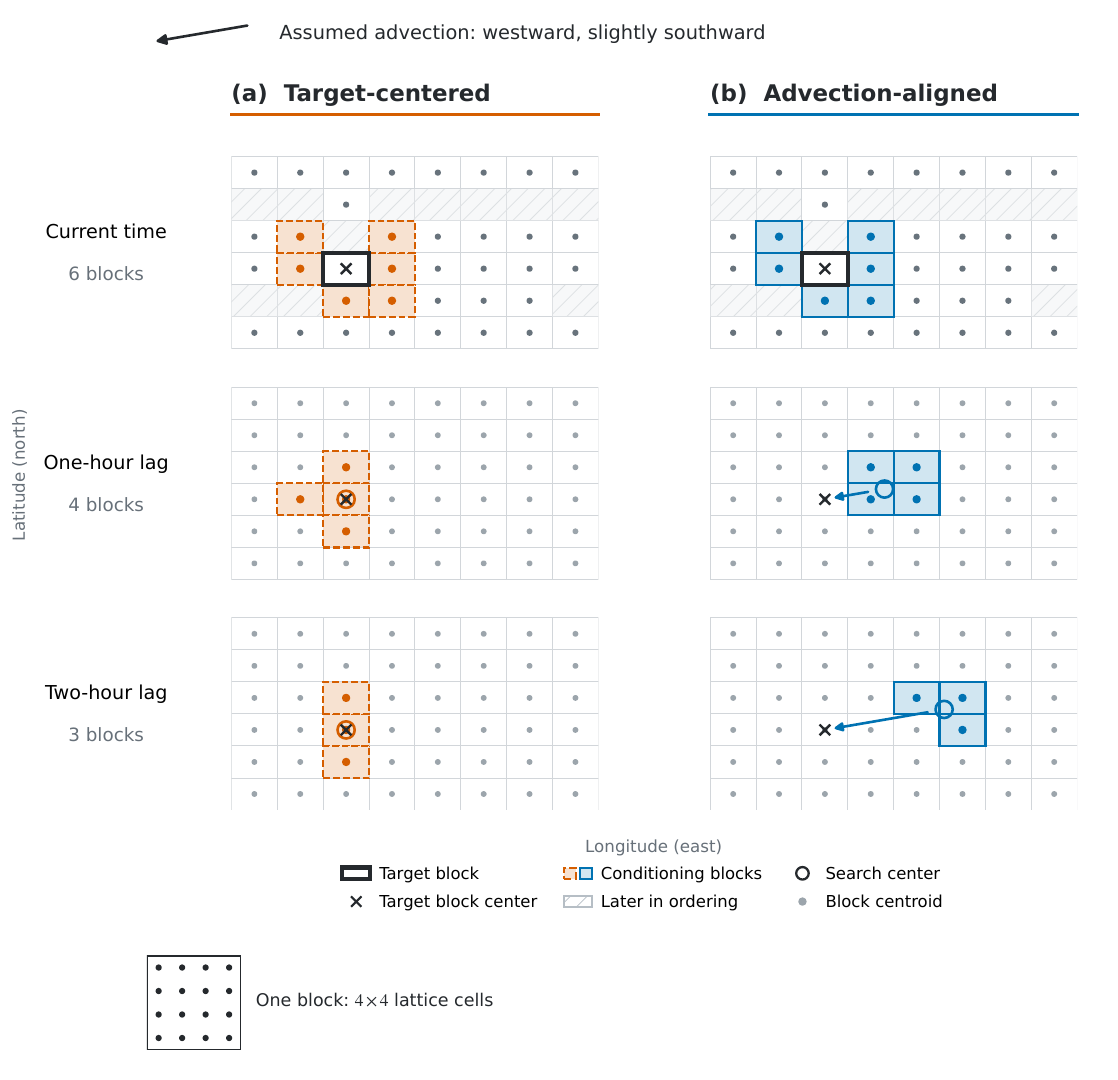}
\caption{{Target-centered (a) and advection-aligned (b)
conditioning with \(4\times4\)-cell blocks and a \(6/4/3\) block budget at
the current hour, one-hour lag, and two-hour lag. Under the illustrative
westward and slightly southward transport, aligned past search centers
are shifted opposite to transport, whereas target-centered searches remain
unshifted. Arrows indicate forward transport toward the target location.}}
\label{fig:conditioning-blocks}
\end{figure*}

{Figure~\ref{fig:conditioning-blocks} illustrates the two constructions
at the same conditioning-block budget.}

The block-level conditioning set for target \((t,j)\) is
\begin{align}
\mathcal C_{t,j}
={}&
\{(t,k):k\in\mathcal N_j^{(0)}\}\nonumber\\*
&{}\cup\{(t-1,k):k\in\mathcal N_j^{(1)}\}\nonumber\\*
&{}\cup\{(t-2,k):k\in\mathcal N_j^{(2)}\},
\label{eq:advection-conditioning-set}
\end{align}
where unavailable temporal layers are omitted at the beginning of the series.
Only observed responses within a selected block enter the point-level
conditioning vector.  Missing cells can therefore reduce the number of
conditioning observations without changing the precomputed block geometry.

The resulting blockwise Vecchia approximation is
\begin{equation}
\widetilde p_{\boldsymbol\theta,\boldsymbol\beta}(\boldsymbol y)
=
\prod_{t=1}^{T}
\prod_{\substack{j=1\\|\mathcal B_j\cap\mathcal O_t|>0}}^{J}
p_{\boldsymbol\theta,\boldsymbol\beta}
\left(
\boldsymbol y_{t,j}
\mid
\boldsymbol y_{\mathcal C_{t,j}}
\right).
\label{eq:advection-block-vecchia}
\end{equation}
Every target observation appears in exactly one factor, while a selected
conditioning block contributes all of its observed components. Therefore,
\(m_0,m_1,m_2\) are block budgets, not numbers of individual observations.

\begin{algorithm}[t]
\caption{Advection-aligned block conditioning graph}
\label{alg:advection-vecchia}
\begin{algorithmic}[1]
\Require Reference lattice and observation masks, block shape, reference
vector \(\widetilde{\boldsymbol v}\), time spacing \(\Delta u\), and block
budgets \((m_0,m_1,m_2)\).
\Ensure Static block conditioning sets \(\{\mathcal C_{t,j}\}\).
\State Partition the lattice into blocks and order their centroids by max--min
ordering.
\For{each ordered block \(j\)}
  \State Let \(\mathcal N_j^{(0)}\) contain up to \(m_0\) nearest
  predecessor blocks.
  \State Compute \(\boldsymbol q_{j,1}\) and \(\boldsymbol q_{j,2}\) using
  \eqref{eq:nominal-query-centres}.
  \State Select \(\mathcal N_j^{(1)}\) and \(\mathcal N_j^{(2)}\) by
  \eqref{eq:nominal-temporal-neighbours}.
\EndFor
\For{each time \(t\) and nonempty target block \(j\)}
  \State Form \(\mathcal C_{t,j}\) from the available temporal layers in
  \eqref{eq:advection-conditioning-set}.
  \State Form \(\boldsymbol y_{\mathcal C_{t,j}}\) from the observed responses
  in the selected blocks.
\EndFor
\State Store the resulting indices for use throughout likelihood optimization.
\end{algorithmic}
\end{algorithm}

At each likelihood evaluation, local covariance matrices are recomputed at
the retained source coordinates using the current covariance parameters,
including \(\boldsymbol v\), through the advection-adjusted spatial lag
\(\boldsymbol h-\boldsymbol v u\); the local conditional contributions are
then evaluated using the batched procedure in
Section~\ref{sec:batched-likelihood}. The block shape and budgets
\((m_0,m_1,m_2)\) are implementation choices; those used below are reported
in Sections~\ref{sec:simulation} and~\ref{sec:gems-application}.

\section{Simulation study}
\label{sec:simulation}

We assess advection recovery and fitted Vecchia objectives
at a matched conditioning budget. The main experiment compares Advection-aligned
with Target-centered across covariance regimes.

\subsection{{Simulation design and evaluation}}
\label{sec:simulation-nominal-fixed}

Each scenario contains 120 simulated eight-hour Gaussian fields: four
independent draws for each of 30 daily GEMS sampling templates from July
2024. We use a fine regular space--time lattice for simulation, with fields
generated by three-dimensional FFT-based circulant embedding
\citep[see][]{guinnessfuentes2017}. Simulated observations are extracted
at the points of the fine simulation lattice nearest to the advection-adjusted GEMS source
locations. The resolution of this simulation lattice and the controlled spectral correction are
described in Appendix~\ref{app:simulation-summaries}. The daily templates
retain the irregular source coordinates and observation masks after
one-to-one assignment to the reference lattice. Advection-aligned and Target-centered
use the same observations, masks, mean specification, initializer, covariance
starting values, and optimizer. Both use $4\times4$ target blocks and budget
$(m_0,m_1,m_2)=(6,4,3)$, allowing at most 208 conditioning observations.
Only the past-time search centers differ: Advection-aligned uses the one- and two-hour
transport locations, while Target-centered uses the unshifted target. Both graphs
remain fixed during optimization, and both fits estimate $\boldsymbol v$
freely.

For spatial lag $\boldsymbol h$ and temporal lag $u$, let
\begin{equation}
\begin{aligned}
d_s^2(\boldsymbol h,u;\boldsymbol v)
 &=\sum_{r=1}^2\frac{(h_r-v_ru)^2}{\rho_r^2},\\
d_t(u)&=|u|/\rho_u.
\end{aligned}
\label{eq:simulation-shifted-distances}
\end{equation}
{We generate and fit a Mat\'ern-$1/2$ covariance:}
\begin{align}
{C}(\boldsymbol h,u)
 &=\sigma^2\exp\{-\sqrt{d_s^2+d_t^2}\}.
 \label{eq:simulation-joint-general}
\end{align}
The fitted covariance matrices use the retained source coordinates and
the same advection-adjusted spatial lag $\boldsymbol h-\boldsymbol vu$.

The reference setting reflects the approximate variability, dependence, and
transport scales observed in the GEMS application. It uses $(\sigma^2,\rho_1,\rho_2,\rho_u)=(10,0.2,0.3,2)$ and
$\boldsymbol v=(0.08,-0.20)^\top$ degrees per hour. Short- and long-range
regimes multiply both spatial ranges by $1/2$ and 2; weak, strong, and zero
advection multiply $\boldsymbol v$ by $1/2$, 2, and 0.
Variance, ranges, and advection are estimated under the generating covariance
family.

\begin{figure*}[!t]
\centering
\includegraphics[width=\textwidth]{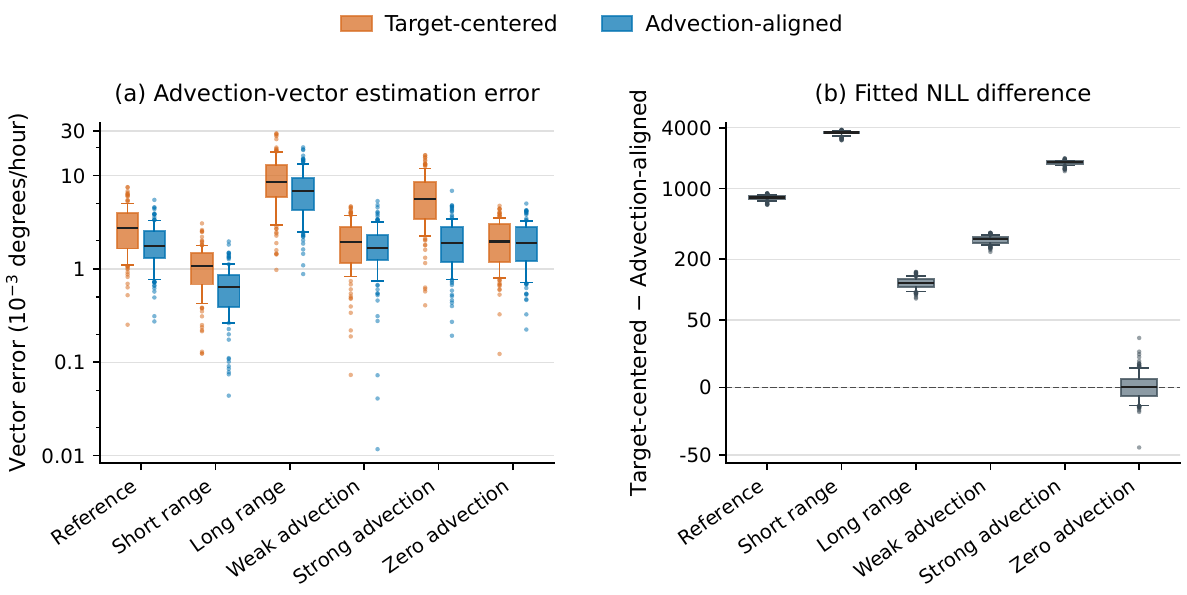}
\caption{{Mat\'ern simulation results from 120 matched
fields per scenario. Boxes show medians and interquartile ranges; whiskers
mark the 10th--90th percentiles, with points beyond. Quantiles are computed
on the original scale. Vertical axes are logarithmic in (a) and symmetric
logarithmic in (b), with a linear region \([-50,50]\).}}
\label{fig:simulation-nominal-fixed}
\end{figure*}

The mean coefficients are profiled by GLS and covariance parameters are
{estimated by L-BFGS with a strong-Wolfe line search
\citep{liunocedal1989,nocedalwright2006}.} A fit is classified as converged when the
maximum absolute gradient of the profiled NLL per target, with respect to the
implemented transformed covariance parameters, is below $10^{-4}$.
We compare advection-vector root mean squared error (RMSE),
\[
\mathrm{RMSE}_v=
\left\{\frac{1}{R}\sum_{r=1}^R
\|\widehat{\boldsymbol v}_r-\boldsymbol v\|_2^2\right\}^{1/2},
\]
and the paired total fitted NLL difference
$\Delta L=L_T(\widehat{\boldsymbol\theta}_T)-
L_A(\widehat{\boldsymbol\theta}_A)$, where each objective uses its own
fixed conditioning graph; $T$ and $A$ denote target-centered and
advection-aligned conditioning. Positive values favor the fitted Advection-aligned density;
they do not establish accuracy relative to the exact likelihood.
Figure~\ref{fig:simulation-nominal-fixed} shows the fieldwise advection-vector
errors $\|\widehat{\boldsymbol v}_r-\boldsymbol v\|_2$ and paired NLL differences.
We also examine variance and range RMSEs. Numerical details and additional
timing summaries
appear in Appendix~\ref{app:simulation-summaries}.

\subsection{{Mat\'ern results}}

{Both methods converge for all 120 fields in each of the six regimes.}
Advection-aligned has lower advection RMSE throughout
({Table~\ref{tab:simready-joint}}), with reductions of
\SimReadyBaselineReduction\% at the reference range,
\SimReadyStrongAdvectionReduction\% under strong advection, and
\SimReadyZeroAdvectionReduction\% under zero advection. The greater gain
under strong transport supports selecting past observations near their
transported locations. At the same $(6,4,3)$ block budget, advection-aligned
also attains a lower fitted NLL in every paired field in the
five nonzero-advection regimes. Mean total NLL differences range
from 117 (long range) to 3,558 (short range); under zero advection the
mean difference is 0.44.

Advection-aligned conditioning reduces the RMSEs of the signal variance and
all three range parameters by about 8\% at the reference range
\reviewblue{(Table~{\hypersetup{linkcolor=black}\ref{tab:simulation-variance-range-rmse}})}. In the
short-range regime, variance,
latitude-range, longitude-range, and temporal-range RMSEs decrease by
9.7\%, 9.4\%, 10.0\%, and 17.5\%, respectively. Under strong advection,
the corresponding reductions are 15.0\%, 13.9\%, 14.3\%, and 37.6\%.
In the long-range, weak-advection, and zero-advection regimes, the
corresponding RMSEs differ by less than 1\% between the two methods.

\begin{table}[!t]
\centering
\begingroup
\color{black}
\caption{\reviewblue{Signal-variance and range RMSE reductions (\%) from 120
matched Mat\'ern fields per regime at budget $(6,4,3)$. Entries are
$100(1-\mathrm{RMSE}_A/\mathrm{RMSE}_T)$, where $A$ and $T$ denote
advection-aligned and target-centered conditioning. Positive values favor
advection-aligned conditioning; negative values favor target-centered
conditioning.}}
\label{tab:simulation-variance-range-rmse}
\small
\setlength{\tabcolsep}{4pt}
\begin{tabular}{@{}lrrrr@{}}
\toprule
Regime & $\sigma^2$ & $\rho_1$ & $\rho_2$ & $\rho_u$ \\
\midrule
Reference range & 7.97 & 7.94 & 8.57 & 7.80 \\
Short range & 9.66 & 9.40 & 9.97 & 17.47 \\
Long range & $-0.12$ & 0.02 & 0.00 & $-0.03$ \\
Weak advection & 0.69 & 0.18 & 0.11 & $-0.20$ \\
Strong advection & 14.98 & 13.92 & 14.32 & 37.62 \\
Zero advection & $-0.85$ & $-0.96$ & $-0.82$ & $-0.48$ \\
\bottomrule
\end{tabular}
\endgroup
\end{table}

{In the reference-, short-, and long-range settings, neither
method produced an advection-vector error above \(0.05^\circ/\mathrm{h}\).}

\FloatBarrier

\section{Application to GEMS total column ozone}
\label{sec:gems-application}

\subsection{{Model and comparison design}}

{We compare advection-aligned and target-centered conditioning on
the 61 daily datasets described in Section~\ref{sec:gems-data}. Each retained
day contains 124,802--144,693 observed targets, and every retained observation
appears in exactly one target block.}

{Exploratory daily fits indicate predominantly westward transport
and generally longer longitudinal than latitudinal correlation ranges.
This westward tendency is qualitatively consistent in direction with the
tropical easterly trade winds \citep{noaa2025tradewinds}.
These features motivate an advection-adjusted spatial lag and separate range
parameters for the latitude and longitude directions.}
For spatial lag \(\boldsymbol h_{ij}\) and temporal lag \(u_{ij}\), we use
{a generalized Cauchy covariance:}
\begin{align}
d_{ij}^2(\boldsymbol v)
&=\sum_{r=1}^2\frac{(h_{ij,r}-v_ru_{ij})^2}{\rho_r^2}
  +\frac{u_{ij}^2}{\rho_u^2},\nonumber\\
C_{\boldsymbol\theta}(i,j)
&=\sigma^2\{1+d_{ij}(\boldsymbol v)^a\}^{-b/a}.
\label{eq:gems-covariance}
\end{align}
The shape parameters
are fixed at \((a,b)=(1,0.5)\) for 2024 and \((0.85,0.55)\) for 2025,
using the same values for the two graphs within each year.

{The Mat\'ern family, including its exponential special case,
provides a standard starting point for spatial covariance modeling
\citep{rasmussenwilliams2006}. In our exploratory analyses, however, fitted
Mat\'ern models underrepresented the observed low-frequency spatial
variability. This motivated the generalized Cauchy family, which allows
separate control of local roughness and algebraic long-distance correlation
decay \citep{gneitingschlather2004}.
\reviewblue{The generalized Cauchy family offers less flexibility in smoothness
than the Mat\'ern family. Our exploratory fits suggested rough ozone fields,
so this limitation did not appear important here.}}

\reviewblue{We used the same year-specific working shape parameters for both
conditioning methods. The values were chosen through exploratory fixed-grid
comparisons and alternating optimization, based on summed daily Vecchia
NLLs within each year. Details of the selection procedure are provided in
Appendix~{\hypersetup{linkcolor=black}\ref{app:gems-implementation}}.}

We estimate six
covariance parameters: \(\sigma^2\), \((\rho_1,\rho_2,\rho_u)\),
and \((v_1,v_2)\). The independent point nugget is fixed at zero.
A diagonal jitter of \(10^{-6}\) is used for numerical stability.

{To accommodate hour-of-day variation, the mean includes an
intercept and seven hourly indicators, with the first observation hour as
the reference. The mean also includes a centered-latitude covariate to model
a linear trend in the mean across latitude, giving nine coefficients in
total. These coefficients are profiled by generalized least squares at each
covariance evaluation. Both conditioning methods use the same mean design.
\reviewblue{Latitude-binned means showed an approximately linear northward
increase (Figure~{\hypersetup{linkcolor=black}\ref{fig:gems-spatial-mean-profiles}},
Appendix~{\hypersetup{linkcolor=black}\ref{app:gems-spatial-mean-profiles}}).
We use a linear latitude term to represent this trend and
hour-specific intercepts to account for diurnal variation over the eight
observation hours.}}

{The two conditioning methods use the same observations, masks,
GLS design, \reviewblue{\(4{\times}4\)}-cell blocks, max--min ordering, covariance family,
initial starting values, and optimizer. Both use the
\((m_0,m_1,m_2)=(6,4,3)\) conditioning-block budget, allowing at most 208
conditioning observations. The only graph difference is the past-time
query center, as defined in Section~\ref{sec:advection-conditioning}.
Each graph is constructed once per day. Both fits initially start from the shared
empirical initializer and estimate covariance advection freely.}

{Computation uses double precision on one NVIDIA L40S GPU
(48 GB nominal memory), processing local systems in chunks of at most 128
target blocks. With fixed block shape and conditioning budget, the number
of local factors increases with the number of target blocks, while the
maximum local matrix dimension remains 224. Implementation and timing
details are provided in Appendix~\ref{app:gems-implementation}.}

Let \(L_{d,T}\) and \(L_{d,A}\) be the fitted, constant-free total Vecchia
NLLs on day \(d\), under the target-centered and advection-aligned graphs, respectively. We
report the paired difference \(\Delta_d=L_{d,T}-L_{d,A}\), so that positive
values favor advection-aligned conditioning. Each total is the NLL
per target multiplied by the actual number \(n_d\) of observed targets.
The omitted Gaussian constant is identical within a pair and cancels.
We report monthly means, medians and interquartile ranges across days separately by year, as well
as \(\sum_d\Delta_d/\sum_dn_d\) to show the scale per observation.
\reviewblue{For each method, we report the Vecchia NLL
evaluated at its fitted parameters using that method's conditioning sets.
A lower fitted NLL does not by itself establish a closer approximation to
the full Gaussian likelihood.}

\subsection{Results}

{Advection-aligned conditioning attains a lower mean daily NLL
in both years and a lower fitted NLL on 53 of 61 days
(Table~\ref{tab:real-data-native-nll}). \reviewblue{Target-centered conditioning was only
slightly better on the days when it achieved a lower fitted NLL.} The two ties occur on days with a
zero empirical initializer, for which the graph constructions coincide.}

\begin{reviewmovedfloat}
\begin{table*}[!t]
\centering
\caption{Matched comparison of fitted daily Vecchia objectives over 61 days. A positive
\(\Delta_d=L_{d,T}-L_{d,A}\) favors advection-aligned conditioning. Mean NLLs are
daily totals with the Gaussian constant omitted. Quartiles describe variation
across dates, not confidence intervals.}
\label{tab:real-data-native-nll}
\small
\setlength{\tabcolsep}{5pt}
\begin{tabular}{@{}lrrrrc@{}}
\toprule
 & \multicolumn{2}{c}{Mean daily total NLL} & & & \\
Period (days) & Target-centered & Advection-aligned & Mean \(\Delta_d\) & Median \(\Delta_d\) [Q1, Q3] & A/T/ties \\
\midrule
July 2024 (31) & 167,861.6 & 167,722.2 & 139.4 & 36.4 [7.7, 192.8] & 26/4/1 \\
July 2025 (30) & 187,987.4 & 187,878.0 & 109.4 & 71.7 [19.9, 89.1] & 27/2/1 \\
Overall (61) & 177,759.6 & 177,634.9 & 124.6 & 46.1 [12.5, 111.0] & 53/6/2 \\
\bottomrule
\end{tabular}
\par\smallskip
\footnotesize A/T/ties denotes the number of advection-aligned wins, target-centered wins,
and ties, using a numerical tie threshold of \(10^{-6}\) in total NLL.
All entries use the same observed targets, masks, mean specification, and
\(6/4/3\) conditioning-block budget within each daily pair.
\end{table*}
\end{reviewmovedfloat}

\begin{reviewmovedfloat}
\begin{figure*}[!t]
\centering
\includegraphics[width=\textwidth]{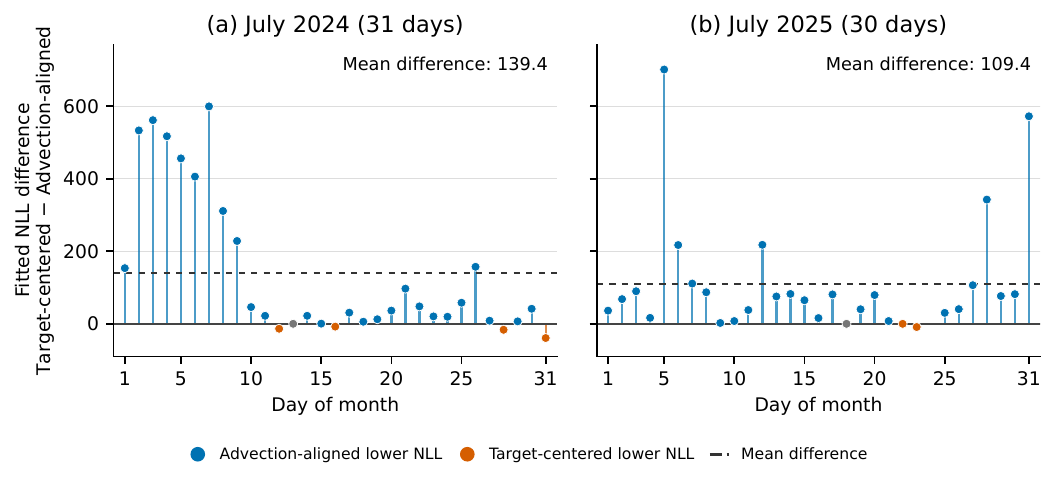}
\color{black}
\caption{Daily differences in fitted total Vecchia NLL, with
each NLL evaluated under its own conditioning graph. Gray points denote
ties.}
\label{fig:gems-nominal-fixed}
\end{figure*}
\end{reviewmovedfloat}

{Daily gains vary across dates
(Figure~\ref{fig:gems-nominal-fixed}). Larger gains early in July 2024
contribute to the gap between its mean and median, whereas the two
summaries are closer in 2025. The aggregate NLL gain per observation is
\(9.86\times10^{-4}\) in 2024 and \(7.83\times10^{-4}\) in 2025.}

\begin{figure*}[!t]
\centering
\includegraphics[width=\textwidth]{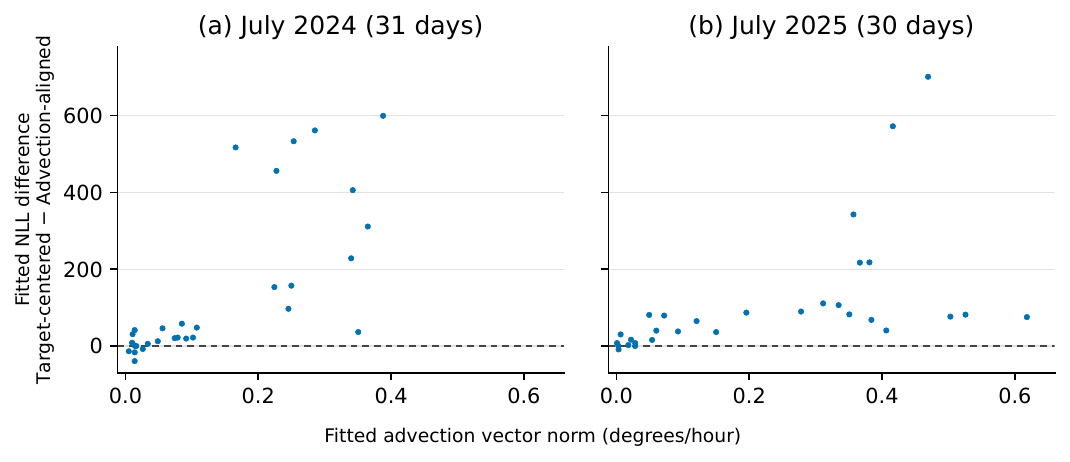}
\caption{{Daily fitted NLL difference versus the norm of the
final advection-aligned estimate.}}
\label{fig:gems-fitted-advection-norm}
\end{figure*}

Figure~\ref{fig:gems-fitted-advection-norm} relates these differences to
\(\|\widehat{\boldsymbol v}_{d,A}\|_2\), the norm of the final
advection-aligned estimate. Larger fitted norms tend to accompany larger
NLL gains, but the relationship is not monotone and differs between years.

Both conditioning rules required about 0.40 seconds per likelihood evaluation.

\FloatBarrier

\section{Discussion}\label{sec:discussion}\label{sec:conclusion}

{Advection-aligned conditioning makes transport geometry explicit
through lag-specific past-time query centers, while both methods fit the
same covariance model with freely estimated advection. Correlation-based
selection can also reflect transport when its covariance includes advection
\citep{kang2023correlation}. Our graph is constructed once and held fixed
during optimization, avoiding repeated neighbor searches.}

{The simulation results show that transport-aligned conditioning can improve
parameter estimation at the same conditioning-block budget.}

{Advection-aligned conditioning attains higher fitted Vecchia
likelihood on 53 of 61 GEMS days, controlling masks, mean specification,
covariance model, optimizer, and block budget, at essentially unchanged
per-evaluation cost. The empirical initializer
requires no preliminary covariance likelihood fit and guides neighbor
selection without constraining the fitted covariance advection parameter.
It also provides an initial advection vector that can be used by other
estimation and neighbor-selection methods.}

\appendix
\setcounter{table}{0}
\renewcommand{\thetable}{\Alph{section}\arabic{table}}
\renewcommand{\theHtable}{appendix.\Alph{section}.\arabic{table}}
\section{{Simulation details}}
\label{app:simulation-summaries}

\begingroup
\makeatletter
\renewcommand{\fnum@table}{{\bfseries\tablename\space\thetable}}
\makeatother
\SimReadyJointTable
\endgroup

{Gaussian fields are generated by FFT circulant embedding on a
regular lattice refined relative to the reference lattice by factors of
100 in latitude and 10 in longitude, with spacings $0.00044^\circ$ and
$0.0063^\circ$, respectively, and one-hour time spacing. The lattice covers
the advection-adjusted GEMS source locations $\boldsymbol s-\boldsymbol vt$
across transport regimes; simulated observations are extracted at the
lattice points nearest to these locations. Nearest-lattice evaluation gives
maximum coordinate discrepancies of $0.00022^\circ$ and
$0.00315^\circ$, respectively. Negative eigenvalues are clipped only if
their spectral-mass fraction is at most $5\times10^{-4}$ and, after
variance renormalization, the relative covariance-distortion bound is at
most $10^{-4}$; otherwise generation fails. Daily templates retain the
closest source observation when multiple observations map to the same
reference cell.}

{The generating mean is $\mu(\boldsymbol s)=260+(s_1+0.5)$.
The fitted mean includes an intercept, centered latitude, and seven hourly
indicators, with coefficients profiled by GLS. All regimes start at
$(\sigma^2,\rho_1,\rho_2,\rho_u)=(10,0.2,0.3,2)$; the same masked-FFT
displacement estimate initializes advection in both methods. Computations
use double precision on an NVIDIA A100 PCIe 40GB GPU in chunks of at most
128 target blocks. Within each scenario, both methods use the same
covariance backend for timing comparisons.}

\section{{GEMS advection estimates}}
\label{app:gems-advection}

\begin{figure*}[!t]
\centering
\includegraphics[width=\textwidth]{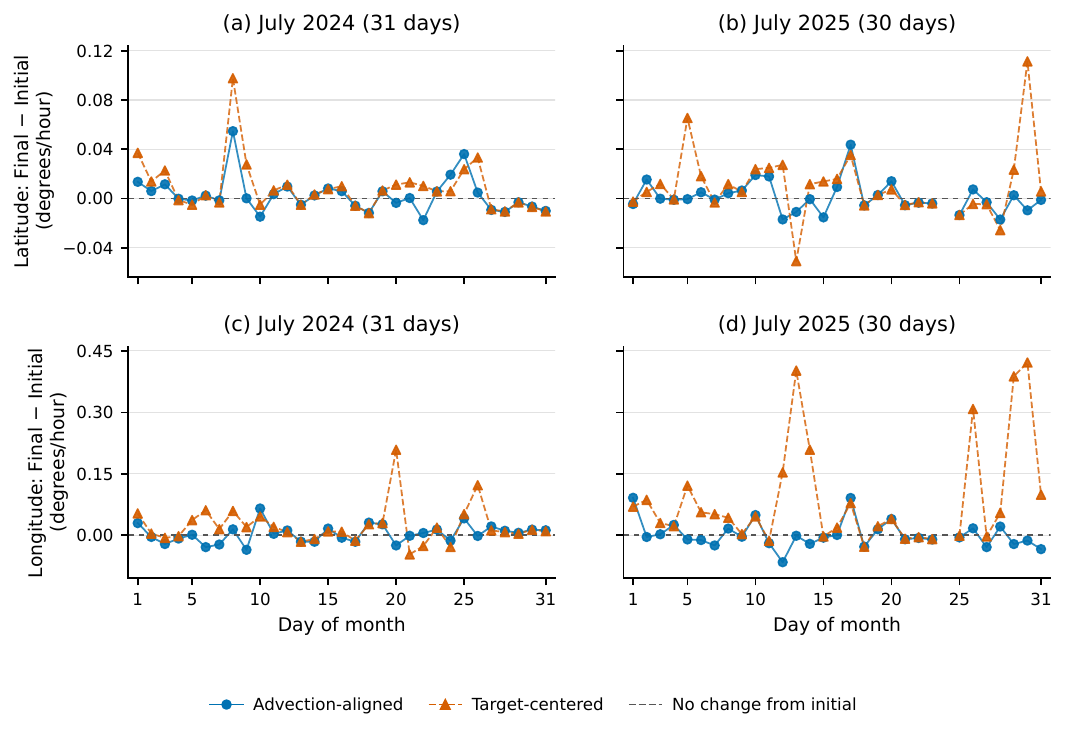}
\caption{{Final advection estimates minus the shared empirical
initializer for the 61 GEMS days. Dashed horizontal lines indicate no change.}}
\label{fig:gems-initial-final-advection}
\end{figure*}

\reviewblue{Figure~{\hypersetup{linkcolor=black}\ref{fig:gems-initial-final-advection}}
compares the empirical advection vector used to construct the advection-aligned
conditioning graph with the final advection estimates from both methods.
The final advection estimates from the advection-aligned fits generally remain
close to this vector, suggesting that the search shifts used for graph
construction are broadly consistent with the fitted advection.}

\section{{GEMS implementation details}}
\label{app:gems-implementation}

{The initializer centers each hourly field by its observed mean
and uses the seven adjacent one-hour transitions. Observations are assigned
to the reference lattice with latitude and longitude spacings
\(0.044^\circ\) and \(0.063^\circ\). Candidate offsets range from \(-20\)
to 20 cells along each axis, giving a \(41\times41\) search grid. We use
\(N_{\min}=1000\), Gaussian smoothing bandwidth
\(h_{\mathrm{smooth}}=0.063^\circ\), and safeguarded quadratic refinement
with Hessian condition-number bound 100. The time spacing is one hour.
Covariance matrices are evaluated at the retained source coordinates.}

{Actual conditioning counts can be smaller because of masks or
unavailable temporal layers. The GPU tensor batch size is distinct from
the conditioning-block budget. Saved hashes verify agreement of the
responses, masks, source coordinates, and mean designs for every pair.}

{The 14 shape-screening combinations use
\(a\in\{0.6,0.7,0.75,0.8,1,1.4,2\}\) at \(b=1\), and
\(a\in\{0.6,0.7,0.75,0.8,1,1.2,1.4\}\) at \(b=0.5\).
Each combination is fitted to the full daily target set for all retained
dates, and combinations are ranked by summed daily total NLL within each
year. These exploratory daily fits allow at most 20 strong-Wolfe L-BFGS
calls with \texttt{max\_iter}=20 and \texttt{max\_eval}=25, history size 40,
and gradient tolerance \(10^{-5}\).}

{The subsequent alternating refinement uses a Powell search
for the two year-shared shape parameters, with at most 60 shape-objective
evaluations per cycle. The six day-specific covariance nuisance parameters
are held fixed during this search and then refitted at the accepted shape
using at most five strong-Wolfe L-BFGS calls with
\texttt{max\_iter}=\texttt{max\_eval}=20, history size 40, and gradient
tolerance \(10^{-5}\); the nine GLS mean coefficients are
recomputed at every likelihood evaluation. Thus, this is alternating
optimization, not full nuisance-profile optimization at every shape trial.
Starting from \((1,0.5)\), the 2024 run reaches
\((1.017233,0.508208)\) after 10 cycles. Starting from \((0.8,0.5)\),
the 2025 run reaches \((0.851516,0.552310)\) after 30 cycles.
Both runs end at their cycle limits; formal convergence or global
optimality is not asserted. The final comparison retains the simple
2024 working values \((1,0.5)\) and uses the rounded 2025 values
\((0.85,0.55)\), rather than the exact refinement endpoints.}

{For the final daily conditioning-method comparison, the
original daily fits use these same L-BFGS settings. Each fit is then
refined from its own saved endpoint using at most 20 additional
strong-Wolfe L-BFGS calls with \texttt{max\_iter}=20,
\texttt{max\_eval}=40, history size 40, and gradient
tolerance \(10^{-5}\) for the NLL per target in transformed covariance
coordinates. A relative total-NLL change threshold of \(10^{-9}\) is
checked between optimizer calls, with the absolute PyTorch
\texttt{tolerance\_change} set to zero.}

{The gradient stopping criterion was met by all 61
advection-aligned fits and all 61 target-centered fits.
All 61 daily pairs are included in the comparison.}

{Computations use 12 CPU threads, PyTorch 2.5.1, and CUDA 12.4.
Graph construction and optimization are timed separately, with GPU
synchronization. Single-evaluation benchmarks use each method's fitted
parameters; likelihood-only timings are medians of three repeated
evaluations.}

\section{{Masked-FFT implementation details}}
\label{app:masked-fft-implementation}

{For a reference lattice of size \(n_1\times n_2\), let
\(N_{\mathrm{grid}}=n_1n_2\). The zero-padded FFT sizes satisfy
\(L_1\geq2n_1-1\) and \(L_2\geq2n_2-1\).}

{The local quadratic refinement in
Section~\ref{sec:advection-initialization} is accepted only when the minimum
has an interior \(3\times3\) stencil with finite values, the fitted Hessian
is positive definite with condition number at most \(\kappa_{\max}\)
in lattice-cell coordinates, and
the fitted stationary point lies within one half-cell in each coordinate
direction; otherwise, the discrete minimum is retained.}

\FloatBarrier
\section{\reviewblue{GEMS spatial mean profiles}}
\label{app:gems-spatial-mean-profiles}

\begin{figure*}[!t]
\centering
\includegraphics[width=\textwidth]{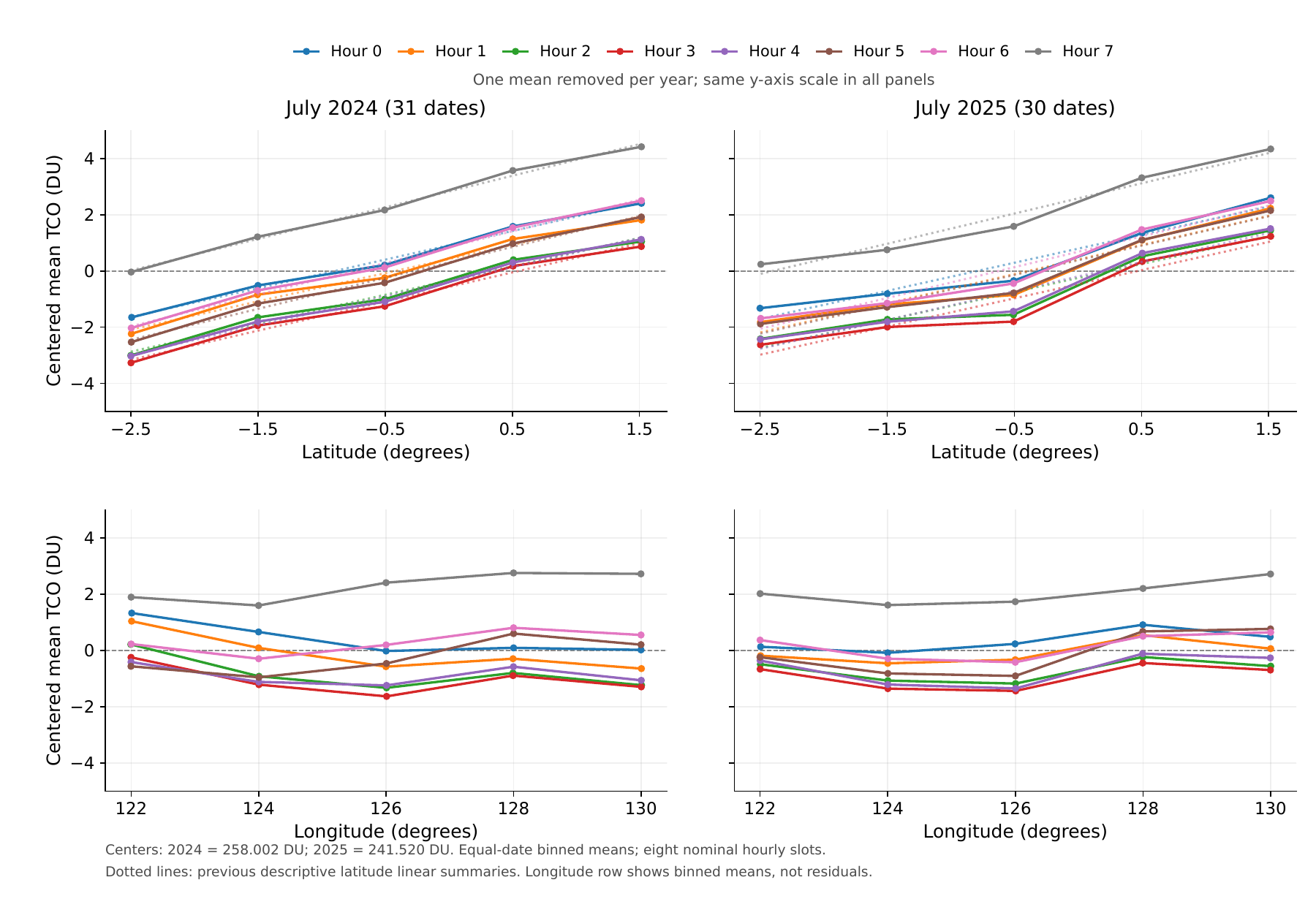}
\caption{\reviewblue{Latitude- (top) and longitude-binned (bottom) total column ozone (TCO) means
for the 31 analyzed days in July 2024 and 30 days in July 2025. Means are
reported in Dobson units (DU) and
calculated within each date, observation hour, and spatial bin, then averaged
equally over dates. One overall mean is removed per year (258.002 DU in 2024;
241.520 DU in 2025), preserving spatial and hourly differences. Colors identify
the eight observation hours. Dotted lines in the upper panels are descriptive
linear fits to each hourly latitude profile, not fitted daily GLS means.}}
\label{fig:gems-spatial-mean-profiles}
\end{figure*}

\FloatBarrier
\backmatter

\section*{Declarations}

\begin{itemize}
\item \textbf{Funding:} No funding was received for conducting this study.
\item \textbf{Conflict of interest:} The authors declare that they have no conflict of interest.
\item \textbf{Data availability:} We acknowledge the Geostationary Environment
Monitoring Spectrometer (GEMS) as the source of the total column ozone data
used in this study. \reviewblue{The data are available from
\url{https://nesc.nier.go.kr}.}
\item \textbf{Code availability:} {The core implementation and synthetic examples for the proposed method will be made publicly available at \url{https://github.com/cl20813/GEMS_TCO_public} upon publication.} The C++ max--min ordering implementation
was adapted from the \texttt{veccs} project maintained by the Katzfuss group
(\url{https://github.com/katzfuss-group/veccs}).
\item \textbf{Author contributions:} Joonwon Lee carried out the analyses
and prepared the manuscript. Michael L. Stein supervised the study.
\end{itemize}

\makeatletter
\immediate\write\@auxout{\string\citation{snMathphysSettings}}
\makeatother
\bibliography{references}


\end{document}